\documentclass[lettersize,journal]{IEEEtran}
\usepackage{amsmath,amsfonts}
\usepackage{algorithmic}
\usepackage{algorithm}
\usepackage{array}
\usepackage[caption=false,font=normalsize,labelfont=sf,textfont=sf]{subfig}
\usepackage{textcomp}
\usepackage{stfloats}
\usepackage{url}
\usepackage{verbatim}
\usepackage{tcolorbox}
\usepackage{graphicx}
\usepackage{cite}
\usepackage{flushend}
\begin{document}
\bstctlcite{IEEEexample:BSTcontrol}
\title{Movable Access Points in Visible Light Communications: Opportunities, Challenges and Future Directions}

\author{Sylvester Aboagye and Telex M. N. Ngatched
\thanks{This paper has been accepted for publication by IEEE. Copyright may be transferred without notice, after which this version may no longer be accessible. }
  }



\maketitle

\begin{abstract}
Visible light communication (VLC) is expected to be a key component of future wireless networks due to its abundant license-free spectrum, inherent high-level security, and the already deployed lighting infrastructure. VLC performance, however, depends on device orientation and the availability of an unobstructed line-of-sight (LoS) link, with transmitter semi-angle and receiver field-of-view (FoV) further affecting alignment, coverage, and reliability. Reconfigurable intelligent surfaces (RISs) can mitigate blockages, orientation issues, and mobility challenges, but their data rates remain far below those of direct LoS links. This article introduces the novel concept of movable access points (MAPs)-aided VLC systems, where dynamically repositioned APs provide new degrees of freedom to ensure LoS connectivity, and transmitter-receiver alignment while providing ultra-high data rates for mobile users. Simulation results show MAPs outperform RIS-aided, fixed-AP, and RIS-only VLC systems in dynamic environments. The article also outlines key challenges and future research directions, including integration with emerging wireless technologies.

\end{abstract}

\section{Introduction}
The advent of fifth-generation (5G) and the evolution toward beyond-5G and sixth-generation (6G) networks are transforming industries and societies, enabling a fully digital and intelligent world. These advances rely on wireless technologies that deliver ultra-reliable, low-latency, and high data rate connections for diverse applications. The impact is most evident indoors, where people spend most of their time, and sectors such as healthcare, education, manufacturing, and entertainment increasingly depend on data-intensive services including real-time health monitoring, remote learning, automation, and immersive media. However, conventional radio frequency (RF) systems struggle to meet these demands due to spectrum congestion, high power consumption, deployment cost, electromagnetic interference (EMI), and security vulnerabilities in dense environments. Visible light communication (VLC) offers a promising alternative by exploiting the abundant unlicensed visible spectrum and widely deployed light-emitting diodes (LEDs), enabling EMI-free, secure, energy-efficient, low-cost, and ultra-high data rate wireless networks \cite{9968053}.

Despite its potential, VLC faces several barriers to widespread adoption, including (i) line-of-sight (LoS) dependency, (ii) vulnerability to blockages and device orientation, (iii) limited mobility support, (iv) short range, (v) narrow transceiver field of view (FoV), (vi) standardization gaps, (vii) bidirectional link support, and (viii) low market readiness. This article focuses on challenges (i)–(v), which have attracted growing attention. Conventional LED-based VLC access points (APs), fixed in position, cannot guarantee LoS coverage for arbitrarily located users. Reconfigurable intelligent surfaces (RISs) have emerged as a low-cost, low-power solution to relax LoS requirements, mitigate blockages and orientation issues \cite{9968053,10586952,9614037}, extend coverage, and address limited FoV \cite{10323414,9354893}. RISs enable non-LoS links by reflecting incident beams toward receivers. However, these links suffer from reflection losses, FoV alignment constraints, and double pathloss, yielding significantly lower data rates than direct LoS. Therefore, ensuring LoS connectivity remains crucial for universal high data rates indoor VLC. 

Movable antenna technology has recently emerged as a cost-effective means to enhance wireless performance \cite{10286328}. By using mechanical controllers, antennas can be repositioned within a confined region to exploit spatial degrees of freedom (DoFs) and improve channel conditions. While extensively studied in RF systems, movable antennas have seen little exploration in optical wireless communications, with only one prior study \cite{10811945}. Importantly, RF and VLC differ significantly in transceiver design, signal properties, and propagation. RF systems employ antennas, whereas VLC relies on LEDs and photodetectors (PDs). LEDs have limited FoV determined by their half-power semi-angle, follow the Lambertian emission model, are ceiling-mounted, and support only short-range LoS propagation while also serving illumination needs. Moreover, the purpose of movable antennas in RF is not the same as that in VLC. In VLC, LEDs may be moved to overcome link blockage and device orientation to provide coverage and establish LoS paths, support mobile users, provide differentiated quality-of-service (QoS), enable beam focusing, and improve security. Furthermore, although both RF movable antennas and VLC movable APs (MAPs) involve mechanical movement, RF systems require only sub-wavelength displacements, whereas VLC necessitates macro-scale track-based movement. This makes VLC MAP design a fundamentally distinct solution. Hence, existing studies (e.g., \cite{10286328,10643473}) on movable antennas for RF systems are not directly applicable to VLC.  To this end, this article first introduces the architecture and DoFs of MAPs in VLC systems. Then, their performance gains in terms of user mobility support, dynamic coverage adaptation, interference mitigation, QoS optimization and link reliability as well as design and implementation challenges are discussed. The achievable data rate of MAP-aided VLC is compared with RIS-aided and fixed-AP VLC systems for different transmit power, varying number of blockers, and mobile users. Finally, technical challenges for future research are discussed.

\section{Fundamentals of MAPs in VLC}
\subsection{Architecture of MAP}
\begin{figure}
 \centering
{{\includegraphics[width=0.45\textwidth]{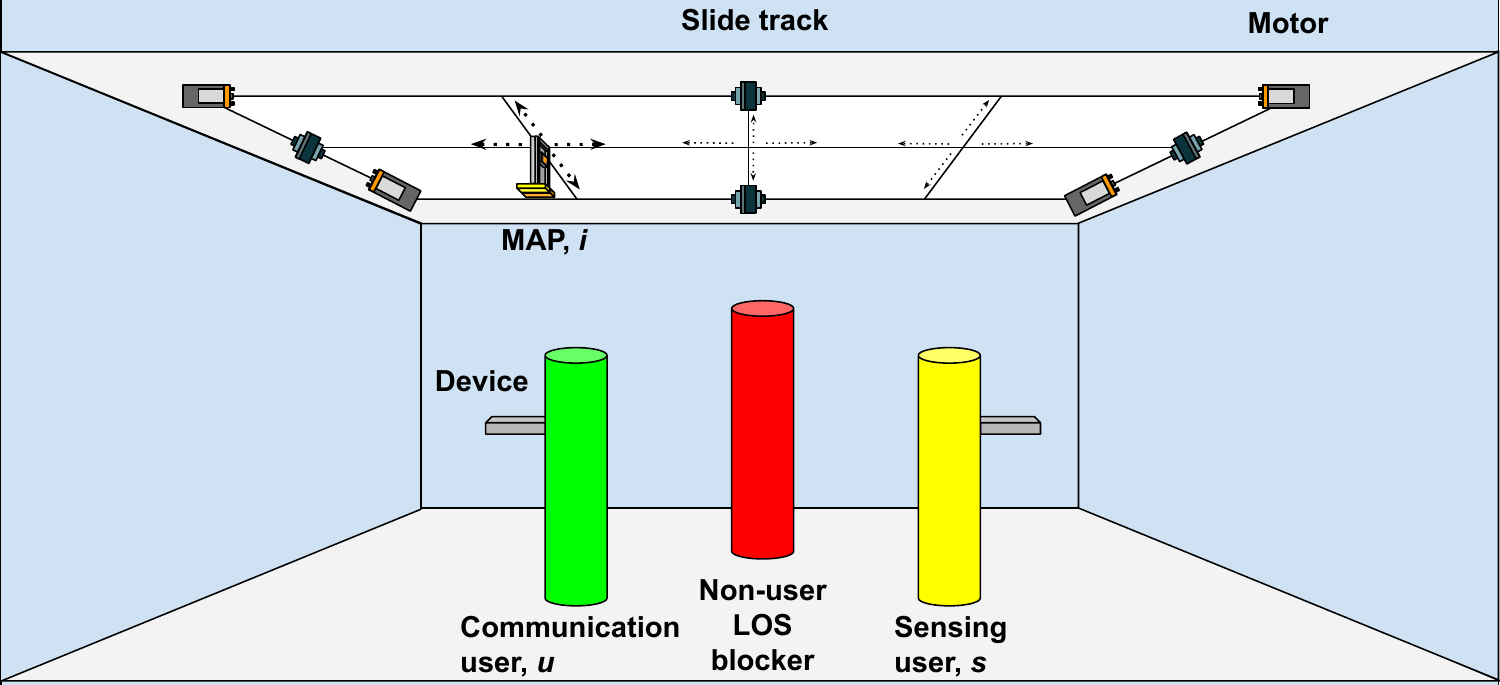}}}
   \caption{\small{MAP-aided indoor VLC system with communication and sensing users and non-user blockers.}}%
 \label{MAP_arc}
 \vspace*{-2mm}
\end{figure}

Figure~\ref{MAP_arc} illustrates an example architecture for a MAP in an indoor VLC system. Mounted on the ceiling, the MAP serves mobile communication and sensing users in the presence of blockers. A MAP comprises a positioning module and a communication module. The positioning module adapts motorized track lighting, widely used in residential and commercial spaces, to enable dynamic adjustments in direction, angle, and beam properties. It consists of: (i) a mechanical subsystem with track rails for fixture movement and power delivery, housing for the LED array, heat management, and motorized sliding components; (ii) an electrical subsystem with conductive strips, flexible wiring, and sliding connectors for continuous power; and (iii) a control subsystem with microcontrollers, motors (e.g., step/servo), and sensors for collision avoidance. The communication module facilitates data and control signal exchange between MAPs, users, and the central processing unit (CPU). The CPU, deployed in the cloud or locally, executes the MAPs' positioning algorithm such as Algorithm~1 of \cite{10811945} using channel state information (CSI), obtainable via uplink wireless fidelity, to determine optimal positions based on user demand and coverage.  In addition, the CPU is responsible for managing network resources such as bandwidth, power, and user-MAP association. Then, it broadcasts the MAPs' optimal positioning and resource allocation as control signals via wired (e.g., power line communication) or wireless communication technology to the control system. Upon receiving these commands, motors relocate the MAP to the target position with the required accuracy.

\subsection{Degrees of Freedom}
\subsubsection{AP Position}
The position of APs in a VLC system strongly affects communication, sensing, and illumination performance. Factors such as AP–user distance, obstacles, and incidence angle directly impact signal-to-noise ratio (SNR) and signal-to-interference-and-noise ratio (SINR), leading to lower SNR/SINR values and reduced data rates. Uneven or poorly planned AP deployment can create non-uniform illumination where users experience significantly reduced data rates, service disruptions, or even complete outage at cell edges. Fixed APs are limited in handling these issues, especially in dynamic environments with mobile users or changing service needs, where variations in channel quality degrade overall performance. MAPs introduce a transformative DoF by enabling real-time reconfigurability in AP positioning. This adaptability allows MAPs to dynamically optimize their locations and adjust cell boundaries in response to user mobility, environmental obstacles, or varying user demands  to maintain ubiquitous high-quality wireless connections.

\subsubsection{AP Orientation}
MAPs offer an effective solution to the device orientation problem in VLC systems. Conventional APs are mounted facing downward to maximize LoS coverage, but performance strongly depends on the interaction between transmitter semi-angle and receiver FoV. Smaller semi-angles confine high-intensity beams, while narrow FoVs increase sensitivity to misalignment, often causing signal loss during orientation changes. Larger semi-angles and wider FoVs mitigate this by broadening coverage and tolerance to device tilts, yet at the cost of higher power consumption, weaker edge performance, and greater inter-cell interference. To simplify analysis, many studies assume users hold devices vertically toward ceiling APs, an unrealistic scenario. MAPs overcome this limitation by dynamically adjusting AP orientation, allowing the light beam to align with receiver orientation and maximize channel gain without sacrificing efficiency.
\subsubsection{MAP Intensity}
MAP intensity, defined as the spatial distribution or density of MAPs within a given area measured in number of MAPs per square meter, is another key DoF in the design, optimization, and operation of MAP-aided VLC systems. Unlike fixed APs, the mobility of MAPs allows dynamic adjustment of intensity based on user density, traffic, and environment. A higher MAP intensity, achieved by deploying more MAPs in a specific area, enhances coverage and reduces the probability of dead zones, ensuring consistent QoS for users in high-density regions. However, increasing MAP intensity introduces challenges, particularly in multi-AP setups, where overlapping light beams can lead to lower SINR, negatively impacting system performance. In addition to MAP intensity, the number of simultaneously active MAPs offers another DoF for interference and energy management. The optimal trade-off between deployment density, number of active MAPs, and system performance requires advanced resource management algorithms.
\subsubsection{Number of MAPs}
In conventional VLC systems, the number of APs is fixed at deployment, based on factors such as illumination requirements, communication coverage goals, expected user density, QoS for various services, room dimensions and geometry, semi-angle of APs, FoV of receivers, cost constraints, and energy efficiency (EE) goals. While optimized for a given space, such systems lack the adaptability to dynamic user density, mobility patterns, or evolving wireless services QoS requirements. MAPs offer a more scalable and promising alternative to fixed designs, where the number of APs can be adjusted in response to real-time user and traffic patterns and changes in service requirements. Their modular track-based design further simplifies reconfiguration: tracks act as both power and communication backbones, enabling MAPs to be added, removed, or repositioned without rewiring or structural changes. This non-invasive flexibility and scalable design makes MAP systems more adaptable to changing user demands, offering a significant advantage over fixed VLC systems.

\subsubsection{Track Rail Architecture}
Track rail design in MAP-enabled VLC systems provides an additional DoF for performance optimization. Layouts such as linear, T-, U-, circular, and grid-based configurations determine AP spatial distribution and should be optimized to align with the room's dimensions, geometry, and user density patterns. For instance, grid-based tracks provide greater flexibility for AP positioning, enabling precise coverage and adaptability, while linear tracks are more suited for corridors or narrow spaces. Additionally, different track designs may support varying power and data transmission capacities, necessitating optimization of the track rail's electrical and communication capabilities to handle a more significant number of MAPs without overloading the system. Track rails can also be installed on ceilings, walls, adjustable mounts, or a combination of these, offering further flexibility in deployment. This versatility allows the strategic placement of MAPs to enhance coverage, improve signal quality, and maximize the overall performance of MAP-aided VLC systems.

\subsection{Feasibility of MAPs in VLC}
The proposed MAP-aided VLC system is readily implementable using commercially available motorized track lighting fixtures, which are already deployed in retail and residential environments. These fixtures integrate compact direct current motors and actuators for movement and orientation, while VLC functionality can be embedded directly into the LED luminaires. Since the motion mechanism consumes only a small fraction of the power relative to LED illumination and communication, and repositioning is infrequent, the added energy overhead is dominated by LED's energy consumption. Moreover, track lighting infrastructure is inherently modular, allowing easy scaling by adding segments or fixtures. This demonstrates that MAP-aided VLC networks can be realized with today’s low-cost and robust hardware with minimal modifications to current lighting installations.

 \begin{figure*}
 \centering
     {{\includegraphics[width=0.9\textwidth]{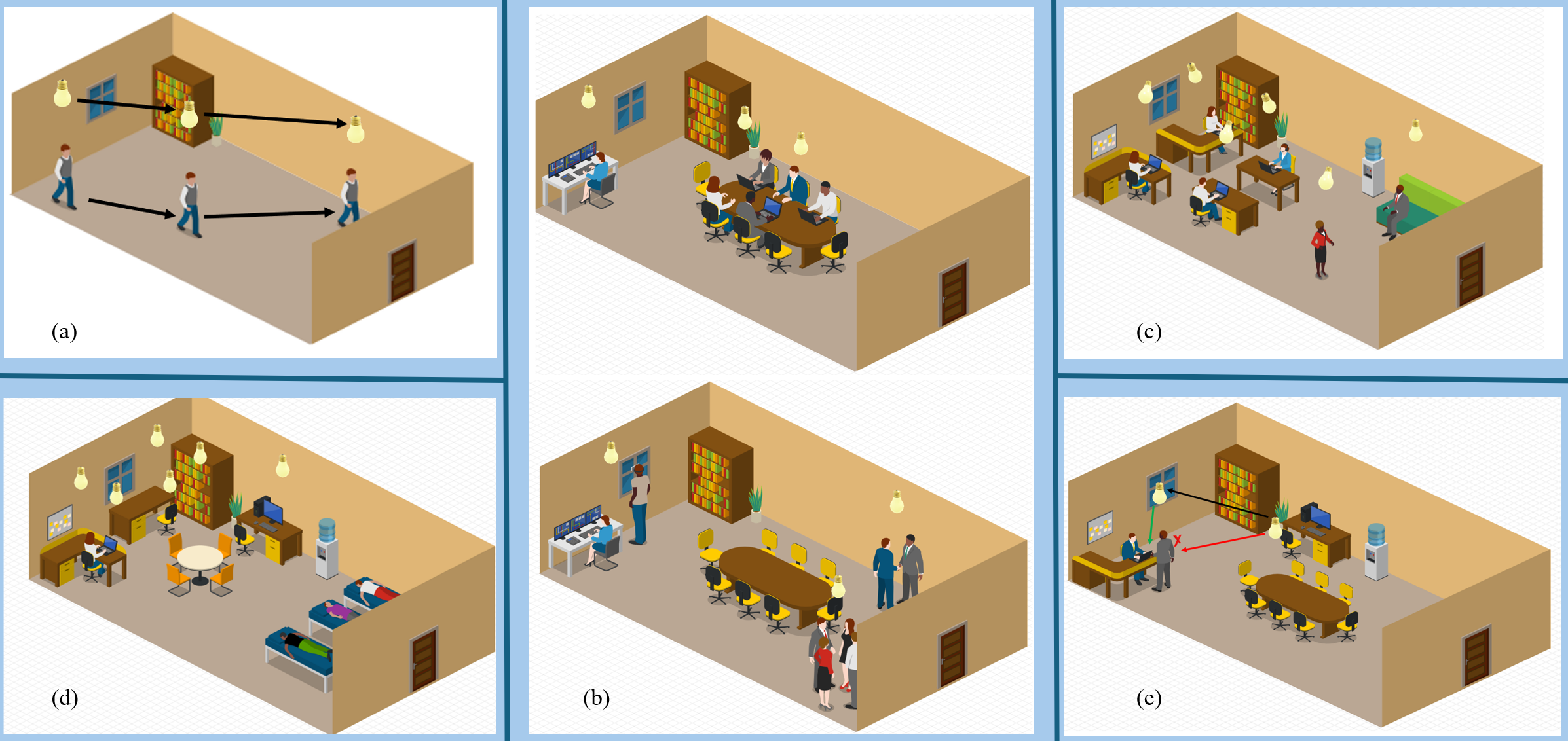} }}%
     \caption{Performance advantages of MAP-aided VLC: (a) Enhanced user mobility support; (b) Dynamic coverage adaptation; (c) Interference mitigation; (d) Quality of service optimization; (e) Enhanced reliability.}%
      \vspace*{-3mm}
   \label{advan}
 \end{figure*}

\section{Performance Advantages of MAPs}
\subsection{Enhanced User Mobility Support}
Unlike fixed APs that have limited flexibility in adapting to user movement, MAPs can move dynamically to maintain an optimal LoS connection with mobile users as shown in Fig.~\ref{advan} (a) for a single user moving from one corner to the other indoors. This capability reduces handover among APs and reduces the likelihood of link degradation or disconnection caused by user movement or changes in device orientation. Moreover, MAPs can proactively adjust their locations to minimize unnecessary handovers, improving user experience and ensuring seamless connectivity. By following user trajectories, MAPs enhance mobility support, making them particularly advantageous in environments with high user dynamics, such as shopping malls, airports, or office spaces.

 \vspace*{-3mm}
\subsection{Dynamic Coverage Adaptation}
One of the key performance advantages of MAPs is their ability to adapt coverage dynamically to changing user densities and patterns. In environments where user distributions are uneven or change frequently as depicted in Fig.~\ref{advan} (b), such as during events or peak office hours, MAPs can relocate to areas with higher user demand to provide targeted coverage while effectively minimizing the concept of cell edge users. For instance, the top part of Fig.~\ref{advan} (b) shows MAP/user arrangment at peak hour while the bottom part shows the arrangement during break hour. This adaptability ensures that resources are allocated efficiently, avoiding underutilization in low-density areas or congestion in high-density zones. Additionally, MAPs can adjust their beam angles, positions, or orientations to optimize coverage for specific regions, ensuring consistent QoS even in complex indoor layouts.

 \vspace*{-3mm}
\subsection{Interference Mitigation}
MAPs play a pivotal role in addressing inter-cell interference, a prevalent challenge in VLC systems with multiple APs. By leveraging their inherent DoFs, MAPs can dynamically adjust their position, orientation, and beam parameters to minimize overlap between adjacent APs and reduce interference as shown in Fig.~\ref{advan} (c) for multiple users and APs scenario. These capabilities not only reduce signal degradation but also enhance overall system performance and user experience in multi-APs environments and complement traditional interference management techniques such as power control and spectrum allocation.
 
  \vspace*{-3mm}
\subsection{Quality of Service Optimization}
The reconfigurability of MAPs allows for more precise optimization of QoS metrics, such as throughput, latency, illumination, and secrecy capacity, to offer a differentiated quality-of-experience for users as illustrated in Fig.~\ref{advan} (d), where some users are working while others are sleeping. In a network with changing key performance metrics, MAPs can dynamically adapt their position, orientation, and intensity to optimize desired metrics at different times. In hybrid RF/VLC systems, MAPs can assist in offloading traffic from congested RF networks, ensuring that users experience consistent performance even in high-demand scenarios. Additionally, MAPs can prioritize certain users or applications by repositioning to maximize channel gain for those requiring high QoS, such as video conferencing or real-time gaming.

 \vspace*{-3mm}
\subsection{Enhanced Reliability}
In scenarios where obstacles or user movements cause link blockages, MAPs can reposition themselves to re-establish LoS links as shown in Fig.~\ref{advan} (e), minimizing service disruptions. This adaptability is particularly valuable in dynamic environments, such as warehouses or retail stores. Moreover, the modular nature of MAP systems allows for quick scalability and redundancy, ensuring continued operation even if individual APs fail. The ability of MAPs to adjust to environmental changes and maintain robust connections contributes to a more resilient and dependable communication network.

\vspace*{-3mm}
\section{Design and Implementation Challenges}\label{DIC}
\subsection{Performance Optimization}
The performance gains of MAPs over fixed APs stem from new DoFs that serve as decision variables in system design. Optimizing MAP parameters -- such as position, orientation, track layout, path planning, and number of MAPs -- for communication, illumination, and sensing remains a key challenge. To maximize the benefits of MAPs, a unified optimization framework that jointly optimizes MAPs design parameters, MAPs-to-users association, and resource allocation is required. This framework must be capable of (i) adapting to user mobility patterns, spatial traffic demands, and environmental conditions; (ii) balancing performance with computational complexity as the number of users and MAPs increases; (iii) accounting for multiple objectives like maximizing coverage and ensuring EE, which often competes with one another. In \cite{10811945}, an optimization framework to jointly optimize MAPs' position and user association for aggregate sum rate and sensing mutual information was investigated. Due to the non-convexity of the resulting optimization problem, a low-complexity algorithm that obtains a locally optimal solution is proposed by leveraging the difference of convex functions and the majorization-minimization technique. This optimization framework can be extended to include the joint optimization of several MAPs' DoFs and resource allocation under various track architectures and connectivity modes (single- and multi-connectivity) to improve its performance further. However, more efficient and low-complexity algorithms capable of handling many decision variables, constraints, and CSI are important problems for further studies. Moreover, unlike existing optimization frameworks for fixed AP VLC systems, positioning delays from motorized MAP movement should be introduced as a constraint or an objective function to balance the trade-off between transmission delays and network performance. 
\vspace*{-3mm}
\subsection{Power Consumption Modeling}
In fixed AP-based VLC, power consumption consists of static and dynamic components. Static power, which remains constant regardless of traffic, includes the operation of LED driver circuits, data processing, compensation for LED inefficiency, and modulation overhead \cite{7437374}. Dynamic power, in contrast, is tied to data transmission activities and fluctuates based on user demand. MAPs introduce new considerations in power consumption modeling. Unlike fixed APs, they can be repositioned or removed when not in use, enabling selective operation that reduces unnecessary fixed energy expenditure in low-traffic areas. However, movement requires additional motion power, and this power depends on the relocation frequency and distance, number of MAPs, trajectory, speed, and motor efficiency. Although motion power is typically negligible compared to transmission power (e.g., LightRail grow light movers consume approximately 5 W for motion \cite{LightRail3P}, whereas LED grow lights typically consume around 500 W), its inclusion is essential for accurate EE analysis, especially in dense deployments. Therefore, new power consumption models are needed to capture mechanical energy costs alongside illumination, communication, and sensing. Intelligent repositioning algorithms can further minimize motion power by predicting user locations and traffic patterns, ensuring that movement occurs only when performance gains justify the energy cost. Overall, the flexibility of MAPs introduces a dynamic component to power consumption modeling, offering the potential for significant EE improvements over fixed APs when trade-offs between motion cost and performance are carefully optimized.

\vspace*{-2mm}
\subsection{Channel Estimation}
Accurate channel estimation is a crucial challenge in MAP-enabled VLC systems, as it directly affects the optimization of MAPs positioning and orientation to maximize system performance. Unlike conventional fixed AP-based systems, which require CSI between fixed APs and static/mobile users, MAPs systems must account for dynamic changes in CSI due to the continuous repositioning and reorientation of APs. This involves mapping the relationship between APs and users' position, their orientation, and the channel gain, significantly increasing the complexity of channel acquisition. In addition, VLC channels are highly sensitive to environmental factors such as light obstructions, reflections, and user mobility, which further complicate the channel estimation process. Conventional channel estimation methods such as APs broadcasting pilots to users and feedback to the APs may fall short in such dynamic settings, requiring novel solutions such as machine learning-based predictive models or real-time adaptive algorithms. Moreover, the need to acquire and process CSI in real-time places a substantial computational overhead and energy consumption burden on the system, especially indoor environments with dense MAPs and placement locations (i.e., high-resolution) due to the need for repositioning MAPs in all potential locations for CSI estimation. More efficient instantaneous/statistical channel acquisition strategies must balance estimation accuracy with computational, energy consumption, and latency constraints, making channel estimation an important challenge in the design and practical implementation of MAP-enabled VLC systems.

\vspace*{-2mm}
\subsection{Practical Constraint}
The deployment of MAPs indoors introduces several practical constraints that must be addressed to ensure efficient and reliable operation. Mechanical limitations, such as finite speed and precision of motors and actuators, can lead to positioning delays and wear over time, affecting system reliability and increasing maintenance needs. Environmental factors, including space limitations, ceiling height, and ambient light, may hinder optimal track rail installation. Moreover, existing infrastructure may require disruptive modifications to accommodate MAPs effectively. In addition, the design of MAPs must align with aesthetic requirements while maintaining structural integrity. For instance, track rails and APs must blend seamlessly into the architectural design without appearing intrusive. Additionally, the weight and distribution of MAP components on ceilings or walls must not compromise structural safety. Achieving this balance between functionality and visual appeal requires innovative approaches. Balancing functionality, reliability, and design is essential to fully realize the potential of MAP-based VLC systems.

\vspace*{-2mm}
\subsection{Backhaul-link Availability}
Backhaul network is an integral part of indoor VLC systems and its quality has an  impact on the overall network performance.  The idea of exploiting existing electrical wiring within buildings for the purpose of backhauling, namely, power line communication, was initially proposed in \cite{1205458}, and has received large attention in the design of fixed-AP VLC systems. Another widely investigated options for backhualing in fixed-AP VLC systems are power-over-ethernet (PoE) \cite{5610962} and optical fiber \cite{7470256}. These wired backhauling approaches can be used in MAP-based VLC systems, but further studies will be required to ensure compatibility with PLC and PoE technologies. Recently, wireless optical  backhauling  (free space optics, visible light, and infrared communications) has attracted significant attention in VLC systems \cite{10706224}, and can be considered a promising option in MAP-based VLC systems  since its deployment would not affect the track rail design. However, employing wireless optical backhaul introduces new requirements for MAPs to have transceivers capable of establishing inter-MAP links, backhaul-aware resource allocation and interference management. Regardless of the the type, all backhauls must be cost-effective and designed to avoid becoming capacity bottlenecks, while maintaining the modularity and flexibility of MAP deployments.

\vspace*{-3mm}
\section{Case Study}
To evaluate the effectiveness of MAPs in overcoming typical challenges in VLC systems such as link blockages, device orientation, and impact of user mobility, we consider the downlink of four indoor VLC environments: 1) MAP-aided VLC system; 2) RIS-aided VLC system; 3) Fixed AP VLC system; and 4) RIS-only VLC system. In the RIS-aided system, data transmission occurs via both direct LoS and RIS-assisted paths. The fixed AP VLC system has no RIS and the LED array is deployed at the room center. In the RIS-only case, no direct LoS path exists. A room size of $10$ m $\times$ $10$ m $\times$ $3$ m, one LED array as the AP, and a single mobile user with movement speed  randomly selected according to a uniform distribution from the range $0.5$ m/s to $2$ m/s are considered. The user and non-users are modeled as cylinders with $0.30$ m diameter and $1.65$ m height. The receiver is held by the user at a distance of $0.75$ m above ground and $0.36$ m from the human body. The orientation of the user and device is random. For the RIS-aided VLC system, an RIS array is deployed on the four walls of the indoor environment. Each RIS array has  $10 \times 40$ mirrors and the dimensions of each mirror are $0.1$ m $\times$ $0.1$ m. Link blockages are considered for the AP-user and RIS-user links. The sine-cosine algorithm proposed in \cite{9543660} is used to optimize the configuration of the mirror array RISs. For the MAP-aided system, a grid-based ceiling track rail with $100$ grid points (i.e., candidate MAP location) is considered. The positioning of the MAP is optimized to maximize data rate using the  algorithm proposed in  \cite{10811945}. The bandwidth is set to $200$ MHz \cite{9543660}. The values for the semi-angle angle, physical area of photodetector, gain of the optical filter, FoV ,  RIS reflection coefficient, photodetector responsivity, thermal noise, and referactive index are chosen as  $60^{\circ}$, $1$ cm$^{2}$, $1$, $70^{\circ}$, $0.95$, $0.53$, $10^{-21}$ A$^2$/Hz, and $1.5$, respectively. The link blockage and user mobility model are adopted from \cite{10811945}. Unless stated otherwise, the number of blockers is set to $16$ and the transmit power to $1$ W. The results have been average over $500$ instances with each having $10$ time slots. 
 
 \begin{figure}
 \centering
     {{\includegraphics[width=0.33\textwidth]{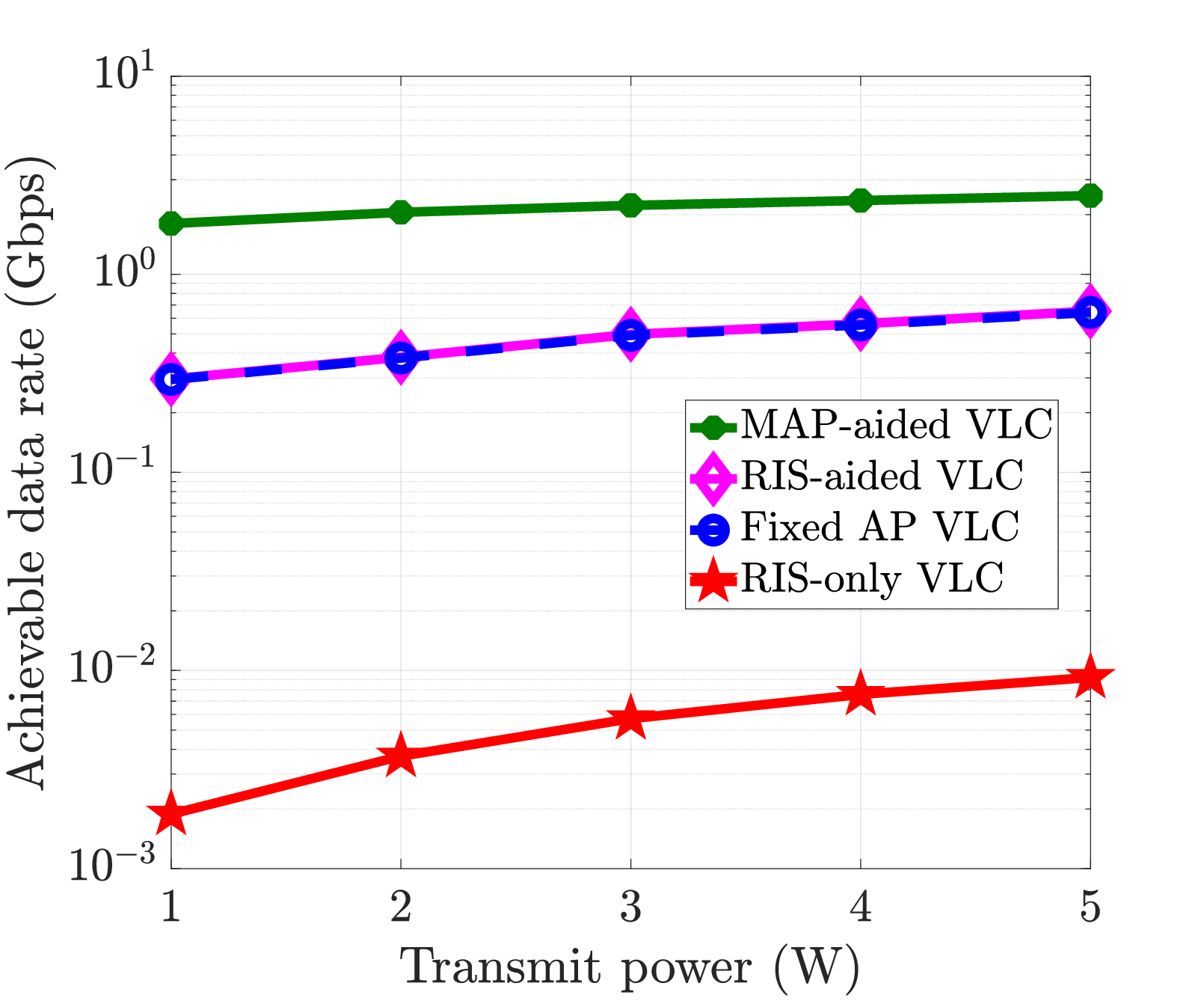} }}%
       \vspace*{-2mm}
     \caption{Achievable data rate vs. transmit power.}%
   \label{sr_fig1}
   \vspace*{-3mm}
 \end{figure}
 
    \begin{figure}
 \centering
     {{\includegraphics[width=0.33\textwidth]{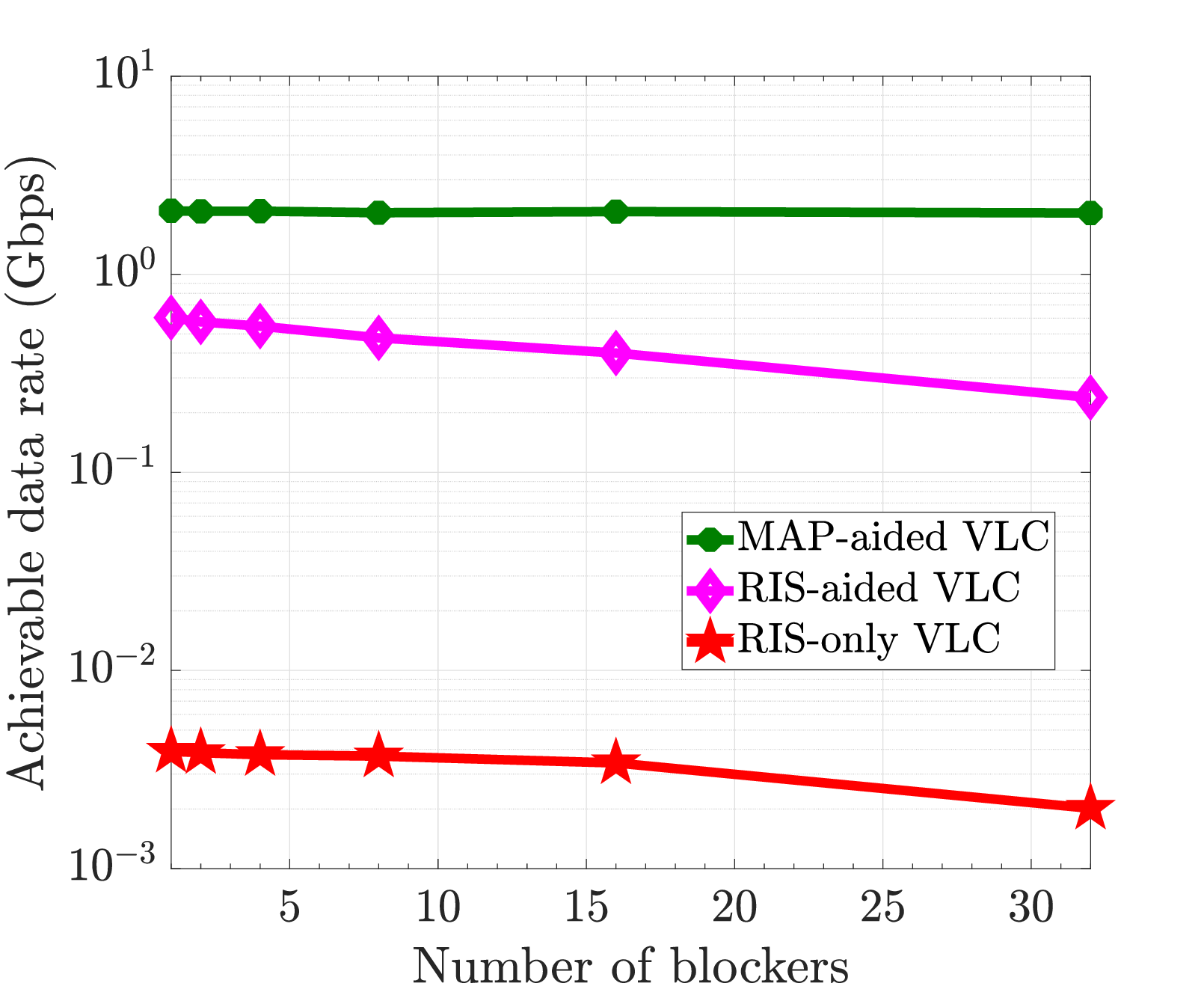} }}%
     \vspace*{-2mm}
     \caption{Achievable data rate vs. number of blockers.}%
   \label{sr_fig2}
   \vspace*{-3mm}
 \end{figure}

Figure~\ref{sr_fig1} illustrates the average achievable data rate of the various system models for different values of the maximum transmit power. It can be observed that the MAP-aided system achieves average performance improvement of $380\%$, $385\%$, and over $50,000\%$ when compared to that of RIS-aided, Fixed AP, and RIS-only VLC systems, respectively. The MAP-aided VLC system achieves the highest data rate since APs are repositioned in real-time to overcome link blockages and device orientation, support user mobility without any handoff cost, and ensure accurate beam alignment such that incidence and irradiance angles are minimized.  The Fixed AP and RIS-aided VLC systems achieve almost similar data rate, with RIS-aided VLC system achieving a marginal gain of $0.1\%$ over Fixed AP.  This is because the data rates of  Fixed AP and RIS-aided VLC systems are largely influenced by LoS connections. It can be seen form Fig.~\ref{sr_fig1} that the data rate for the RIS reflections (i.e., RIS-only VLC) is relatively insignificant when compared to that of the Fixed AP VLC (tens  vs. hundreds of Mbps). This figure confirms the importance of guaranteeing LoS connections in VLC systems as the data rate from non-LoS links such as RIS reflections gets significantly reduced due to double pathloss effects and reflection losses.

Figure~\ref{sr_fig2} reveals the impact of varying the number of blockers on the achievable data rate for MAP-aided, RIS-aided, and RIS-only VLC systems. Fixed AP VLC system is not considered since the acheivable data rate is almost the same as RIS-aided VLC system. The transmit power is set to $2$ W. It can be observed that increasing the number of blockers from $1$ to $32$ leads to an average data rate loss of $2\%$, $153\%$, and $93\%$ for the MAP-aided, RIS-aided, and RIS-only VLC systems, respectively. The data rate gain of MAP-aided VLC system over RIS-aided and RIS-only systems increased from   $246\%$ and  $53,000\%$ to $754\%$ and $101,000\%$, respectively, for $1$ and $32$ blockers. This highlights the effectiveness of MAP-aided VLC in guaranteeing high data rate in dense blockers scenarios.

 \begin{figure}
 \centering
     {{\includegraphics[width=0.33\textwidth]{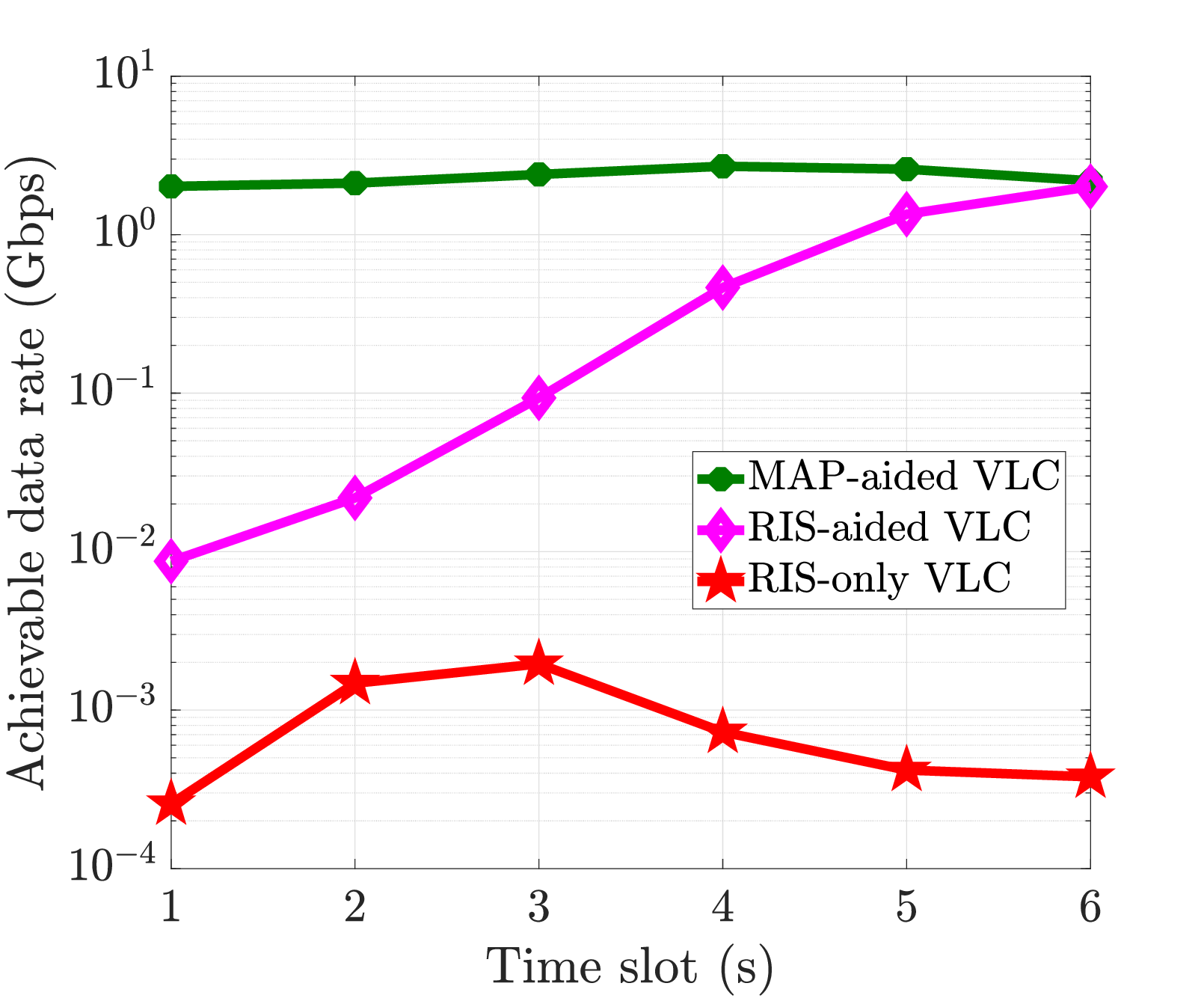} }}%
     \vspace*{-2mm}
     \caption{Achievable data rate vs. time slot.}%
   \label{sr_fig3}
   \vspace*{-3mm}
 \end{figure}
 
   \begin{figure}
 \centering
     {{\includegraphics[width=0.33\textwidth]{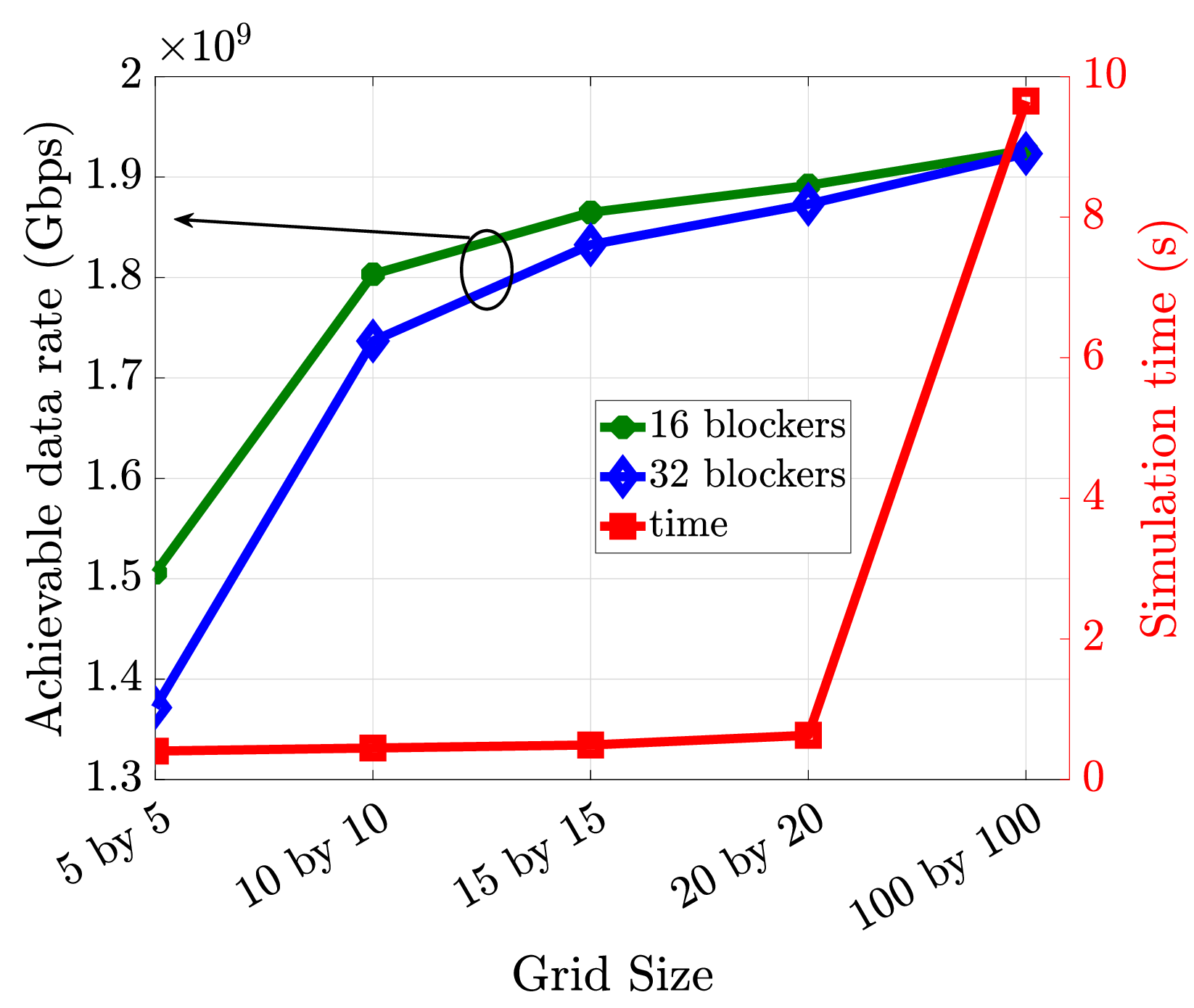}}}%
     \vspace*{-2mm}
     \caption{Achievable data rate vs. grid size.}%
   \label{sr_figl}
   \vspace*{-3mm}
 \end{figure} 

Figure~\ref{sr_fig3} investigates user mobility support in MAP-aided, RIS-aided, and RIS-only VLC systems. Fixed AP VLC system is not considered since the acheivable data rate follows the same trend as RIS-aided VLC system. In this figure, the mobile user begins from a room corner, and moves towards the center. The MAP-aided VLC system consistently maintains ultra-high data rates as the mobile user moves towards the center of the room. This is because of the flexibility of the MAP to adaptively position itself to maintain optimal LoS connection with user. The RIS-aided VLC system starts at a lower data rate compared to MAP-aided VLC but increases steadily as the user moves towards the center. The reason is that the signal strength and probability of LoS connection increase as the user moves to the room center.   At the center of the room, MAP-aided VLC outperforms RIS-aided VLC in terms of data rate  by an average gain of $9\%$.  The RIS-only VLC system exhibits a non-monotonic trend in data rate. Initially, as the user moves away from the corner of the room, the data rate increases because the coverage and signal strength from the two nearest RISs improve, even though the remaining two RISs contribute weaker reflections due to longer distances. However, as the user approaches the center of the room, the link distances from all four RISs become approximately equal and increase simultaneously, leading to higher path loss and a decline in the achievable data rate. This highlights the limitations of an RIS-only VLC system, which lacks a direct LoS link and depends entirely on nearly-passive signal reflection.

Finally, Fig.~\ref{sr_figl} shows the effect of increased grid size on data rate and the trade-off with time complexity. It can be observed that increasing the grid size improves the achievable data rate by offering more flexibility in MAP placement. The most notable gains occur when moving from 5$\times$5 to 10$\times$10, while further increases up to 20$\times$20 provide only small improvements. Beyond this point, performance almost saturates, whereas time complexity grows rapidly, especially at 100$\times$100. Hence, grids between 10$\times$10 and 20$\times$20 provide the best balance between data rate and computational cost.

 \vspace*{-2mm}
\section{Future Directions}In addition to the challenges discussed in Section~\ref{DIC}, this section presents promising research directions for future studies towards the practical realization of MAP-aided VLC systems.

\vspace*{-2mm}
\subsection{Synergy between MAPs and ML}
MAP mobility enables dynamic AP placement and orientation, creating a highly adaptable environment for communication and sensing. Machine learning can exploit this mobility to learn real-time optimal configurations and balance communication, illumination, and sensing objectives in ways traditional optimization methods cannot, especially in practical settings with non-ideal positioning errors and limited MAP speed.

\vspace*{-2mm}
\subsection{RIS-enabled MAP-aided VLC Systems}
Combining RISs with MAPs allows joint dynamic control of the optical channel and AP positioning, opening new opportunities for interference management, coverage optimization, and enhanced security. Future work should focus on low-complexity joint optimization, the choice and deployment of RIS types, and practical deployment considerations such as energy efficiency and scalability in multi-user VLC systems.

\vspace*{-2mm}
\subsection{MAP-assisted Hybrid VLC/RF Systems}
MAPs can actively shape VLC coverage to complement RF links, enabling adaptive hybrid networks that respond to user mobility and environmental changes. Key research challenges include coordinated MAP and RF antenna movement, handover mechanisms, and resource allocation strategies for seamless, energy-efficient operation.

\vspace*{-2mm}
\subsection{Role of RSMA/NOMA in MAP-aided VLC Systems}
MAPs allow dynamic adjustment of AP positions to create distinct channel gains among users, improving the feasibility and performance of rate splitting multiple access (RSMA) and non-orthorgonal multiple access (NOMA). This mobility-driven approach can enhance spectral efficiency, fairness, and physical layer security in ways not possible with fixed APs.

\vspace*{-2mm}
\section{Conclusion}
Compared to fixed APs in VLC, MAPs offer several DoFs to enhance VLC performance through real-time dynamic positioning in confined regions. This article provided a comprehensive overview of MAP-aided VLC systems, an innovative paradigm for ensuring reliable LoS connections for arbitrarily located users. First, the architecture of MAPs and their key DoFs for performance optimization were introduced. Then, their numerous advantages over conventional APs were highlighted. Subsequently, the various design and implementation challenges for MAPs, along with  promising solutions, were examined. A case study was presented to illustrate the siginificant data rate improvement in MAP-aided VLC system over fixed AP and RIS-aided VLC systems. The results confirmed that LoS connectivity is critical for practical VLC deployment as the data rate for RIS non-LoS links is relatively insignificant and may  fail to meet the demands of new wireless applications. For the first time, MAPs demonstrated the ability to guarantee LoS connections for arbitrarily located indoor users. Finally,  the potential integration of MAPs with other emerging wireless technologies and open research were explored, aiming to inspire further work and support MAP adoption in future wireless networks.

\vspace*{-2mm}

\bibliographystyle{IEEEtran}
\bibliography{IEEEfull}

\end{document}